\documentstyle[multicol,pre,aps]{revtex}

\begin{document}
\draft
\title{{\bf On Coupled Directed Percolation Processes: a Unifying View}}
\author{{\sc H.K.\ Janssen}}
\address{Institut f\"{u}r Theoretische Physik III, Heinrich-Heine-Universit\"{a}t,\\
40225 D\"{u}sseldorf, Germany }
\date{\today}
\maketitle

\begin{abstract}
It is shown that the universal critical properties of two recently
introduced coupled directed percolation processes can be described by a
single ra\-pi\-di\-ty reversal invariant stochastic reaction-diffusion
model. It is demonstrated that all renormalizations needed for the
calculation of the universal scaling behavior near the multicritical point
can be gained from the Gribov process (Reggeon field theory). Consequently
the crossover exponent describing the scaling of the linear coupling
parameter is given by $\Phi =1$ to all orders of the perturbation expansion.
\end{abstract}


\begin{multicols}{2}
\narrowtext


\label{Anfang}Nonequilibrium processes, their stationary states and their
phase transitions have been of considerable interest in na\-tu\-ral science as well as
in medicine and sociology since many years. Such processes can often be modelled by
growth and decay of populations with spatially local interaction rules. The transition
between survival and extinction of the population is a nonequilibrium continuous phase
transition phenomenon and is characterized by universal scaling laws. For the
description of transitions in systems that show active and absorbing inactive states,
percolation models play an outstanding role. Some years ago it was conjectured
\cite{Ja81,Gr82} that Markovian growth mo\-dels with one-component order parameters
displaying a transition into an absorbing state in the absence of conservation laws
belong generically to the universality class of directed percolation (DP). Besides DP
\cite{BrHa57,CaSu80,Ob80} this universality class includes e.g. Reggeon field theory
(RFT) \cite{Gribov,Mo78,GrassbGr},\ the contact process \cite{Ha74,Li85}, certain
cellular automata \cite{Ki83} and some catalysis models \cite{ZGB86,GLB89}. It is
relevant to a vast range of models in physics, chemistry, biology and sociology.

For the description of universal behavior near a cri\-ti\-cal transition it is often
useful to model a universality class by mesoscopic stochastic processes involving the
order parameter and other relevant fields. In case of the DP class a representation by
the Langevin equation for the density variable of the so-called Gribov process (the
stochastic version of the Schl\"{o}gl model \cite{Sch72}) is appropriate. The name
Gribov process was coined by Grassberger who showed that RFT is rather a Markov process
in disguise than a quantum theory \cite{GrassbGr}. On the level of a formulation of
stochastic processes by means of path integrals, there is superficially no difference
between RFT and the Gribov process. However RFT uses creation and annihilation
operators for particles as the principal fields in contrast to the densities and their
conjugate response fields of the Gribov process. Thus, the latter one is more adapted
to a mesoscopic description.

We have been studying the Gribov process for some time by renormalized field theoretic
methods \cite{Ja81}. By ge\-ne\-ra\-li\-za\-tion to the multicomponent case we have
shown recently that the Gribov process with many in\-ter\-ac\-ting species (directed
percolation with many colors) belongs to the DP class as well\cite {Ja97}. In the
multicomponent model the evolution of the species is described by a set of Langevin
equations for density variables $\left\{ n_{i}\left( {\bf r},t\right) \right\} $. These
stochastic reaction-diffusion equations are such that the condition of the existence of
an absorbing va\-cu\-um state individually for each species is not destroyed. Therefore
they are at least bilinear coupled in the densities. We have shown that multicolored
directed percolation exhibits a new interesting phenomenon: Even if the dynamics of the
different colors is modelled symmetrically, there exists an instability which leads to
a dif\-fe\-ren\-tia\-tion of the species in their active state under coarse graining.
The universal properties of this color symmetry breaking are described by a
multicomponent Gribov process with in pairs unidirectional coupled species.

Very recently further interest on coupled directed percolation processes emerged
\cite{AEHM96}. T\"{a}uber et al. \cite{THH98} have introduced a model that couples two
species unidirectional (without feedback) by a linear term in the equations of motion.
Such a term induces always a nonzero mean-density of the second species in a state
where the first species (the parent) is active. Thus only the inactive phase of the
first species can be truly absorbing.

The features of the bilinear and the linear unidirectional coupled models
for two species are combined in the Langevin equations
\begin{eqnarray*}
\lambda ^{-1}\frac{\partial n_{1}}{\partial t} &=&\nabla ^{2}n_{1}-\Bigl(
\tau _{1}+\frac{g}{2}n_{1}\Bigr) n_{1}+\zeta _{1}\;, \\
\lambda ^{-1}\frac{\partial n_{2}}{\partial t} &=&\nabla ^{2}n_{2}-\Bigl(
\tau _{2}+\frac{g}{2}n_{2}+fn_{1}\Bigr) n_{2}+\sigma n_{1}+\zeta _{2}\;,
\end{eqnarray*}
\vspace{-0.6cm}
\begin{equation}
\langle \zeta _{i}\left( {\bf r},t\right) \zeta _{j}\left( {\bf r}^{\prime
},t^{\prime }\right) \rangle =\lambda ^{-1}g^{\prime }n_{i}\left( {\bf r} ,t\right)
\delta _{ij}\delta \left( {\bf r-r}^{\prime }\right) \delta \left( t-t^{\prime }\right)
\;.  \label{01}
\end{equation}

Here, for simplicity, equal kinetic coefficients $\lambda $, intraspecies couplings
$g$, and noise parameters $g^{\prime }$ for each species are assumed. As remarked
above, the linear coupling $\propto \sigma $ destroys the absorption condition for
species $2$ in a state where species $1$ is active. Therefore T\"{a}uber et al. argue
that one should introduce further couplings in Eqs.\ (\ref{01}) because they are now
generated through coarse graining. This leads to the completed stochastic
reaction-diffusion equations
\begin{eqnarray}
\lambda ^{-1}\frac{\partial n_{1}}{\partial t} &=&\nabla ^{2}n_{1}-\Bigl(
\tau _{1}+\frac{g}{2}n_{1}\Bigr) n_{1}+\zeta _{1}\;,  \nonumber \\
\lambda ^{-1}\frac{\partial n_{2}}{\partial t} &=&\nabla ^{2}n_{2}-\Bigl(
\tau _{2}+\frac{g}{2}n_{2}\Bigr) n_{2}  \nonumber \\
&&+\Bigl(\sigma -\frac{f_{1}}{2}n_{1}-f_{2}n_{2}\Bigr)n_{1}+\zeta _{2}\;,
\label{02}
\end{eqnarray}
with Gaussian correlations of the Langevin forces with mean zero given by
\begin{eqnarray}
\langle \zeta _{1}\left( {\bf r},t\right) \zeta _{1}\left( {\bf r}^{\prime
},t^{\prime }\right) \rangle &=&\lambda ^{-1}g^{\prime }n_{1}\left( {\bf r} ,t\right)
\delta \left( {\bf r-r}^{\prime }\right) \delta \left( t-t^{\prime }\right) \;,
\nonumber \\
\langle \zeta _{1}\left( {\bf r},t\right) \zeta _{2}\left( {\bf r}^{\prime
},t^{\prime }\right) \rangle &=&\lambda ^{-1}f_{2}^{\prime }n_{1}\left( {\bf
r},t\right) \delta \left( {\bf r-r}^{\prime }\right) \delta \left( t-t^{\prime }\right)
\;,  \nonumber \\
\langle \zeta _{2}\left( {\bf r},t\right) \zeta _{2}\left( {\bf r}^{\prime
},t^{\prime }\right) \rangle &=&\lambda ^{-1}\Bigl(g^{\prime }n_{2}\left( {\bf
r},t\right) +f_{1}^{\prime }n_{1}\left( {\bf r},t\right) \Bigr)
\nonumber \\
&&\qquad \qquad \times \delta \left( {\bf r-r}^{\prime }\right) \delta
\left( t-t^{\prime }\right) \;.  \label{03}
\end{eqnarray}
Some remarks are in order here. We consider the coupled DP processes always in the form
of Gribov processes, i.e.\ in terms of stochastic reaction-diffusion equations for
density variables. These are in general the appropriate mesoscopic variables for the
description of the percolation-type transition between an absorbing inactive phase and
an active one as in our case. This description is legitimate since creation and
annihilation processes allowing for single-particle annihilation lead to strong spatial
clustering. The field theory of T\"{a}uber et al., which as RFT is based on creation
and annihilation operators for particles on a lattice, is, up to irrelevancies, fully
equivalent to our mesoscopic approach.

\label{full}In the following we focus on the scaling behavior of
correlation and response functions in the vicinity of the multicritical point where the
strongly relevant parameters $\tau _{i}$ and $\sigma $ are small. Critical properties
are extracted from the mesoscopic model by using renormalized field theory and the
renormalization group equation in conjunction with an $\varepsilon $-expansion about
the upper critical dimension\cite{Am84,ZiJu93,BJW76,DDP78,Ja79}. It is convenient to
recast the stochastic equations of motion as a dynamic functional \cite
{Ja79,DeDo76,Ja76,Ja92}. The Langevin equations (\ref{02},\ref{03}) lead to the
following dynamic functional in renormalized form
\begin{eqnarray}
{\cal J}_{c} &=&\int dtd^{d}x\lambda  \nonumber \\
\times\!\!\!\!\!\!\!&&\bigg\{\tilde{s}_{1}\Big[Z\lambda ^{-1}\partial
_{t}+Z_{\tau }\tau _{1}-Z_{\lambda }\nabla ^{2}+\frac{g}{2}Z_{g}^{\,}\left(
s_{1}-\tilde{s}_{1}\right) \Big]s_{1}  \nonumber \\
&&+\tilde{s}_{2}\Big[Z\lambda ^{-1}\partial _{t}+Z_{\tau }\tau
_{2}-Z_{\lambda }\nabla ^{2}+\frac{g}{2}Z_{g}^{\,}\left( s_{2}-\tilde{s}
_{2}\right) \Big]s_{2}  \nonumber \\
&&+\tilde{s}_{2}\Big[2AZ\lambda ^{-1}\partial _{t}-Z_{\sigma }\sigma
-Y_{1}\tau _{1}-Y_{2}\tau _{2}-Z_{b}b\nabla ^{2}  \nonumber \\
&&+\frac{f_{1}}{2}Z_{1}s_{1}+f_{2}Z_{2}s_{2}-f_{2}^{\prime }Z_{2}^{\prime }
\tilde{s}_{1}-\frac{f_{1}^{\prime }}{2}Z_{1}^{\prime }\tilde{s}_{2}\Big]s_{1}
\bigg\}\;.  \label{04}
\end{eqnarray}

The $s_{i}\left( {\bf r},t\right) \sim n_{i}\left( {\bf r},t\right) $ are the suitable
rescaled density fields such that $g^{\prime }=g$. The $
\widetilde{s}_{i}\left( {\bf r},t\right) $ denote the response fields
(corresponding to the conjugated au\-xi\-li\-ary variables of the operator formulation
of statistical dynamics by Kawasaki \cite{Kaw71} and Martin, Siggia, and Rose
\cite{MSR72}). Correlation and response functions can now be expressed as functional
averages of monomials of the $s_{i}$ and $\tilde{ s }_{i}$ with weight $\exp \left\{
-{\cal J}_{c}\right\} $. The responses are defined with respect to additional local
particle sources $\widetilde{h}
_{i}\left( {\bf r},t\right) \geq 0$ in the equations of motion (\ref{02}).
These sources lead to the additional terms $-\left( \widetilde{ h}_{1}
\widetilde{s}_{1}+\widetilde{h}_{2}\widetilde{s}_{2}\right) $ in the
integrand of the dynamic functional ${\cal J}_{c}$ (\ref{04}). As usual, the various
$Z$-factors and counterterms $A$, $Y_{i}$ are introduced to absorb the infinities that
are generated in perturbation theory by taking the continuum limit. Using dimensional
re\-gu\-la\-ri\-za\-tion and minimal renormalization the different counterterms arise
as pole expansions in $
\varepsilon .$ Note that for a complete renormalization a further gradient
term with a new dimensionless constant $b$ has to be introduced as well as
further additive renormalizations that are not considered in \cite{THH98}.

The scaling by a suitable mesoscopic length and time scale, $\mu ^{-1}$ and $
\left( \lambda \mu ^{2}\right) ^{-1}$ respectively, leads to $\widetilde{s}
_{\alpha }\sim s_{\alpha }\sim \mu ^{d/2},$ $g\sim f_{i},f_{i}^{\prime }\sim
\mu ^{\varepsilon /2}$ where $\varepsilon =4-d$. Hence $d_{c}=4$ is the
upper critical dimension. A further inspection of Eq.\ (\ref{04}) and a glance at the
perturbation expansion shows that the functional ${\cal J}_{c}$ includes all possible
relevant couplings with respect to the Gaussian fixed point as the starting point of
the perturbation expansion and is thus renormalizable. This expansion is dimensionally
re\-gu\-la\-ri\-zed and minimal renormalized, whereupon the bare (unrenormalized)
quantities are related to the renormalized ones by the scheme
\begin{eqnarray}
\mathaccent"7017{\tilde{s}}_{1} &=&Z^{1/2}\bigl( \tilde{s}_{1}+A\tilde{s}
_{2}\bigr) \;,\qquad \mathaccent"7017s_{2}=Z^{1/2}\bigl(
s_{2}+As_{1}\bigr) \;,  \nonumber \\
\mathaccent"7017\lambda &=&Z^{-1}Z_{\lambda }\lambda \;,\qquad\qquad\quad\;\, \mathaccent
"7017\tau _{i}=Z_{\lambda }^{\,-1}Z_{\tau }\tau _{i}\;,  \nonumber \\
\mathaccent"7017g &=&Z^{-1/2}Z_{\lambda }^{\,-1}Z_{g}^{\,}g\;,\qquad\quad\,
\mathaccent"7017b=Z_{\lambda }^{\,-1}Z_{b}b-2A\;,  \nonumber \\
\mathaccent"7017\sigma &=&Z_{\lambda }^{\,-1}\Bigl(Z_{\sigma }\sigma +\bigl(
Y_{1}+AZ_{\tau }\bigr) \tau _{1}+\bigl( Y_{2}+AZ_{\tau }\bigr) \tau _{2}
\Bigr)\;,  \nonumber \\
\mathaccent"7017f_{1} &=&Z_{\lambda }^{\,-1}Z^{-1/2}\Bigl(
Z_{1}f_{1}-A\left( 1-A\right) Z_{g}^{\,}g-2AZ_{2}f_{2}\Bigr) \;,  \nonumber
\\
\mathaccent"7017f_{1}^{\prime } &=&Z_{\lambda }^{\,-1}Z^{-1/2}\Bigl(
Z_{1}^{\prime }f_{1}^{\prime }-A\left( 1-A\right) Z_{g}^{\,}g-2AZ_{2}^{\prime
}f_{2}^{\prime }\Bigr) \;,  \nonumber \\
\mathaccent"7017f_{2} &=&Z_{\lambda }^{\,-1}Z^{-1/2}\Bigl(
Z_{2}f_{2}-AZ_{g}^{\,}g\Bigr) \;,\qquad  \nonumber \\
\mathaccent"7017f_{2}^{\prime } &=&Z_{\lambda }^{\,-1}Z^{-1/2}\Bigl(
Z_{2}^{\prime }f_{2}^{\prime }-AZ_{g}^{\,}g\Bigr) \;.  \label{05}
\end{eqnarray}
Dimensionless coupling constants are defined as
\begin{equation}
u=G_{\varepsilon }g^{2}\mu ^{-\varepsilon }\;,\qquad v_{i}^{\left( \prime
\right) }=G_{\varepsilon }^{1/2}f_{i}^{\left( \prime \right) }\mu
^{-\varepsilon /2}\;,  \label{06}
\end{equation}
where $G_{\varepsilon }=\Gamma \left( 1+\varepsilon /2\right) /\left( 4\pi
\right) ^{d/2}$. As argued in \cite{Ja97}, the intraspecies renormalization
factors $Z,Z_{\lambda },Z_{\tau },Z_{g}^{\,}$ are functions of $u$ only and
therefore given by the well known renormalizations of the one-species model.

Before we consider the calculation of the various renormalizations to
one-loop order we introduce a linear, homogeneous transformation between the
fields. Let us call it the $\alpha $-transformation:
\begin{eqnarray}
\tilde{s}_{1} &\rightarrow &\tilde{s}_{1}-\alpha \tilde{s}_{2}\;,\qquad
s_{1}\rightarrow s_{1}\;,  \nonumber \\
s_{2} &\rightarrow &s_{2}+\alpha s_{1}\;,\qquad \tilde{s}_{2}\rightarrow
\tilde{s}_{2}\;.  \label{07}
\end{eqnarray}
Note that a corresponding $\beta $-transformation
\begin{eqnarray}
\tilde{s}_{1} &\rightarrow &\tilde{s}_{1}+\beta \tilde{s}_{2}\;,\qquad
s_{1}\rightarrow s_{1}\;,  \nonumber \\
s_{2} &\rightarrow &s_{2}+\beta s_{1}\;,\qquad \tilde{s}_{2}\rightarrow
\tilde{s}_{2}\;,  \label{08}
\end{eqnarray}
is exploited to eliminate $\mathaccent"7017{\tilde{s_{2}}}\partial _{t}
\mathaccent"7017s_{1}$ from the unrenormalized form of the functional ${\cal
J}_{c}$. Likewise the scaling transformation $s_{i}\rightarrow \gamma s_{i}$ ,
$\tilde{s}_{i}\rightarrow \gamma ^{-1}\tilde{s}_{i}$ is employed to specify $g$ as the
coupling constant for the nonlinear term and the noise term of the intra-species parts
of the dynamic functional (\ref{04}). The $
\alpha $-transformation of ${\cal J}_{c}$ does not change the form of this
functional. However it leads to new coupling constants that we denote with
an overbar:
\begin{eqnarray}
\bar{\sigma} &=&\sigma +\alpha \left( \tau _{1}-\tau _{2}\right) \;,
\nonumber \\
\bar{f}_{1} &=&f_{1}+2\alpha f_{2}+\alpha \left( \alpha -1\right) g\;,\qquad
\bar{f}_{2}=f_{2}+\alpha g\;,  \nonumber \\
\bar{f}_{1}^{\prime } &=&f_{1}^{\prime }-2\alpha f_{2}^{\prime }+\alpha
\left( \alpha +1\right) g\;,\qquad \bar{f}_{2}^{\prime }=f_{2}^{\prime
}-\alpha g\;.  \label{09}
\end{eqnarray}
The transformed couplings give rise to transformed renormalization factors
which we denote by overbars also:
\begin{eqnarray}
\bar{A} &=&A\;,\qquad \bar{Z}_{b}=Z_{b}\;,\qquad \bar{Z}_{\sigma }=Z_{\sigma
}\;,  \nonumber \\
\bar{Y}_{1} &=&Y_{1}+\alpha \left( Z_{\tau }-Z_{\sigma }\right)
\;,\!\!\!\qquad \bar{Y}_{2}=Y_{2}-\alpha \left( Z_{\tau }-Z_{\sigma }\right)
\;,  \nonumber \\
\bar{Z}_{1}\bar{f}_{1} &=&Z_{1}f_{1}+2\alpha Z_{2}f_{2}+\alpha \left( \alpha
-1\right) Z_{g}^{\,}g\;,  \nonumber \\
\bar{Z}_{1}^{\prime }\bar{f}_{1}^{\prime } &=&Z_{1}^{\prime }f_{1}^{\prime
}-2\alpha Z_{2}^{\prime }f_{2}^{\prime }+\alpha \left( \alpha +1\right)
Z_{g}^{\,}g\;,  \nonumber \\
\bar{Z}_{2}\bar{f}_{2} &=&Z_{2}f_{2}+\alpha Z_{g}^{\,}g\;,\qquad \bar{Z}
_{2}^{\prime }\bar{f}_{2}^{\prime }=Z_{2}^{\prime }f_{2}^{\prime }-\alpha
Z_{g}^{\,}g\;.  \label{010}
\end{eqnarray}
Via the transformation formulas (\ref{09}), the possible fixed point values of the
coupling constants are transformed as well. Thus, the form invariance of the dynamic
functional under the one-parametric $\alpha $-transformation explains the appearance of
the fixed lines found in \cite{THH98}. Therefore it is appropriate to define $\alpha
$-invariant dimensionless coupling constants as
\begin{eqnarray}
w_{1} &=&u^{1/2}\left( v_{1}+v_{2}\right) -v_{2}^{\,2}\;,  \nonumber \\
w_{1}^{\prime } &=&u^{1/2}\left( v_{1}^{\prime }+v_{2}^{\prime }\right)
-v_{2}^{\prime \,2}\;,  \nonumber \\
w_{2} &=&u^{1/2}\left( v_{2}+v_{2}^{\prime }\right) \;.  \label{011}
\end{eqnarray}

We define rapidity reversal as
\begin{equation}
\tilde{s}_{1}\left( t\right) \leftrightarrow -s_{2}\left( -t\right)
\;,\qquad s_{1}\left( t\right) \leftrightarrow -\tilde{s}_{2}\left(
-t\right) \;.  \label{012}
\end{equation}
It turns out that under this inversion all unprimed quantities in ${\cal J}
_{c}$ (\ref{04}) change to the primed ones and vice versa, and the
``temperature'' parameters $\tau _{1},\tau _{2}$ are exchanged. Furthermore the
renormalization functions are related by
\begin{eqnarray}
Z_{i}\left( \left\{ v\right\} ,\left\{ v^{\prime }\right\} \right)
&=&Z_{i}^{\prime }\left( \left\{ v^{\prime }\right\} ,\left\{ v\right\}
\right) \;,  \nonumber \\
Y_{1}\left( \left\{ v\right\} ,\left\{ v^{\prime }\right\} \right)
&=&Y_{2}\left( \left\{ v^{\prime }\right\} ,\left\{ v\right\} \right) \;,
\label{013}
\end{eqnarray}
whereas $Z_{\sigma },A,Z_{b}$ are invariant under the exchange $\left\{ v\right\}
\leftrightarrow \left\{ v^{\prime }\right\} $. Note that $ Z,Z_{\lambda },Z_{\tau
},Z_{g}$ do not dependent from the sets $\left\{ v\right\} ,\left\{ v^{\prime }\right\}
$ at all. Especially if $\left\{ v\right\} =\left\{ v^{\prime }\right\} $ (a property
that is not broken by renormalization), the functional ${\cal J}_{c}$ is rapidity
reversal invariant. Then we have $Z_{i}=Z_{i}^{\prime }$ and $Y_{1}=Y_{2}$.

The somewhat lengthy but simple calculation of all the one-loop
renormalizations leads to the Wilson functions of the renormalization group
equation. In part they can be gained from the results of \cite{THH98} in the
case $b=0$. If we set $u$ to its fixed point value $u^{\ast }=2\varepsilon
/3+O\left( \varepsilon ^{2}\right) $ and define as usual $\beta
_{p}=\partial p/\partial \ln \mu |_{0}$, where $p$ is any of the coupling
constants and $|_{0}$ denotes a differentiation hol\-ding the unrenormalized parameters
constant, we obtain for the $\beta $-functions of the invariant couplings
\begin{eqnarray}
8\beta _{w_{1}} &=&4\left( 2w_{1}+w_{2}-u^{\ast }\right) w_{2}+4u^{\ast
}w_{1}^{\prime }  \nonumber \\
&&+16u^{\ast }a-u^{\ast }b\left( 8w_{1}+10w_{2}\right) +9\left( u^{\ast
}b\right) ^{2}\;,  \nonumber \\
8\beta _{w_{1}^{\prime }} &=&4\left( 2w_{1}^{\prime }+w_{2}-u^{\ast }\right)
w_{2}+4u^{\ast }w_{1}  \nonumber \\
&&+16u^{\ast }a-u^{\ast }b\left( 8w_{1}^{\prime }+10w_{2}\right) +9\left(
u^{\ast }b\right) ^{2}\;,  \nonumber \\
4\beta _{w_{2}} &=&8\left( w_{2}-u^{\ast }\right) w_{2}+4u^{\ast }\left(
w_{1}+w_{1}^{\prime }\right)   \nonumber \\
&&+8u^{\ast }a-6u^{\ast }bw_{2}+3\left( u^{\ast }b\right) ^{2}\;,  \nonumber
\\
8u^{\ast }\beta _{b} &=&u^{\ast }\left( w_{1}+w_{1}^{\prime }\right) +\left(
w_{2}-u^{\ast }\right) w_{2}  \nonumber \\
&&+16u^{\ast }a-u^{\ast }b\left( w_{2}+u^{\ast }\right) +\left( u^{\ast
}b\right) ^{2}\;.  \label{014}
\end{eqnarray}
The function $a$ results from the additive renormalization $A$ and mixes the
renormalized fields in a correlation or response function under application
of the renormalization group as
\begin{equation}
{\cal D}_{\ln \mu }\left\{ \tilde{s}_{1},s_{1},\tilde{s}_{2},s_{2}\right\}
=\left\{ -a\tilde{s}_{2},0,0,-as_{1}\right\} \;.  \label{015}
\end{equation}
Here ${\cal D}_{\ln \mu }$ is the usual linear differential renormalization
group operator given by
\begin{eqnarray}
{\cal D}_{\ln \mu } &=&\mu \partial _{\mu }+\kappa _{\tau }\tau \partial
_{\tau }+\left( \kappa _{\sigma }\sigma +\kappa _{1}\tau _{1}+\kappa
_{2}\tau _{2}\right) \partial _{\sigma }  \nonumber \\
&&+\zeta \lambda \partial _{\lambda }+\sum_{p}\beta _{p}\partial _{p}+\frac{
\gamma }{2}\;,  \label{016}
\end{eqnarray}
with $\zeta =\gamma -\gamma _{\lambda },$ $\kappa _{\tau }=\gamma _{\lambda
}-\gamma _{\tau },$ $\kappa _{\sigma }=\gamma _{\lambda }-\gamma _{\sigma }$
and $\kappa _{i}=-y_{i}-a$. The $y_{i}$ results from the additive
renormalizations $Y_{i}$ and $\gamma _{..}=\partial \ln Z_{..}/\partial \ln
\mu |_{0}$. The function $a$ is found to be
\begin{eqnarray}
a &=&-\Bigl(2\left( w_{1}+w_{1}^{\prime }\right) +2\left( w_{2}-u\right) w_{2}/u
\nonumber \\ &&-4bw_{2}+3ub^{2}\Bigr)/16\;.  \label{017}
\end{eqnarray}
To one-loop order we have the well known RFT results $\gamma =-u/4,$ $\gamma
_{\lambda }=-u/8,$ $\gamma _{\tau }=-u/2$. The new re\-nor\-ma\-li\-za\-tions yield
\begin{eqnarray}
\gamma _{\sigma } &=&\frac{1}{2}\Bigl( bu-w_{2}\Bigr) \;,  \nonumber \\
8y_{1} &=&4\Bigl(u^{1/2}v_{1}+v_{2}v_{2}^{\prime }\Bigr)  \nonumber \\
&&-6bu^{1/2}v_{2}-2bu^{1/2}v_{2}^{\prime }+3b^{2}u\;,  \nonumber \\
8y_{2} &=&4\Bigl(u^{1/2}v_{1}^{\prime }+v_{2}v_{2}^{\prime }\Bigr)  \nonumber
\\
&&-6bu^{1/2}v_{2}^{\prime }-2bu^{1/2}v_{2}+3b^{2}u\;.  \label{018}
\end{eqnarray}

In order to determine the fixed-point solutions of Eqs.\ (\ref{014}), $\beta
_{p}^{\ast }=0$, we impose the condition $a^{\ast }=b^{\ast }=0$. This yields
$w_{1}^{\ast }=w_{1}^{\prime \ast }=0$, with $w_{2}^{\ast }=0$ (instable) or
$w_{2}^{\ast }=u^{\ast }$ (stable). It can easily be checked that these solutions are
consistent with the full set of Eqs.\ (\ref{014}) and (\ref {017}). These are the
solutions found by T\"{a}uber et al. \cite{THH98}. Note that on the fixed lines
generated from the stable fixed point by the $
\alpha $ -transformation two special points are found: A rapidity reversal
invariant one with $v_{2}^{\ast }=v_{2}^{\prime \ast }=-2v_{1}^{\ast }=-2v_{1}^{\prime
\ast }=\sqrt{u^{\ast }}/2$ and a minimally coupled one with $ v_{1}^{\ast }=v_{1}^{\prime
\ast }=v_{2}^{\prime \ast }=0,\;v_{2}^{\ast }=
\sqrt{u^{\ast } }$. The former is related to the latter through an $\alpha $
-transformation (\ref{07}) with $\alpha =-1/2$.

Now one has to prove stability of the fixed points of the full equations (\ref{014})
without using the constraints $a^{\ast }=b^{\ast }=0$. A linearization about
$w_{1}=w_{1}^{\prime }=b=0$ and either $w_{2}=0$ or $ w_{2}=u^{\ast }$ shows that the
flow of $b$ is instable for $w_{2}=0$, whereas it shows full stability of the fixed
point for $w_{2}=u^{\ast }$. This vindicates the neglect of $a,b$ and the corresponding
counterterms $A$ and $Z_{b}b$ in \cite{THH98} at the stable fixed point. However
without further knowledge this statement is only correct in the one-loop calculation
and could be destroyed in higher loop orders. We will show that the stable fixed point
is given by $w_{1}^{\ast }=w_{1}^{\prime \ast }=0$, $w_{2}^{\ast }=u^{\ast }$ to all
orders of the loop expansion. As a consequence the fixed point values $a^{\ast
},b^{\ast },y_{1}^{\ast },y_{2}^{\ast }$ are zero, and $\gamma _{\sigma }^{\ast
}=\gamma _{\tau }^{\ast }$. This leads to a crossover exponent $\Phi =1$ where $\Phi $
determines the scaling of $\sigma
/|\tau _{i}|^{\Phi }$.

\label{restricted}In the following we will demonstrate that a restricted
model with only one coupling constant comprises all re\-le\-vant universal features and
especially the stable fixed point of all the coupled directed percolation processes. We
revisit the Langevin equations (\ref{01}) without the additional couplings introduced
by T\"{a}uber et al.\ \cite{THH98}. After the suitable rescaling of the densities
$n_{i}$ to the stochastic variables $s_{i}$ we now get the renormalized dynamic
functional
\begin{eqnarray}
{\cal J}_{m} &=&\int dtd^{d}x\,\lambda   \nonumber \\ \times\!\!\!\!\!\!\!&&
\biggl\{\tilde{s}_{1}\Bigl[Z\lambda ^{-1}\partial _{t}+Z_{\tau }\tau _{1}-Z_{\lambda
}\nabla ^{2}+\frac{g}{2}Z_{g}^{\,}\left( s_{1}-\tilde{s }_{1}\right) \Bigr]s_{1}
\nonumber \\ &&+\tilde{s}_{2}\Bigl[Z\lambda ^{-1}\partial _{t}+Z_{\tau }\tau
_{2}-Z_{\lambda }\nabla ^{2}+\frac{g}{2}Z_{g}^{\,}\left( s_{2}-\tilde{s}
_{2}\right) \Bigr]s_{2}  \nonumber \\
&&+\tilde{s}_{2}\Bigl[-Z_{\sigma }\sigma +Z_{f}fs_{2}\Bigr]s_{1}\biggr\}\;.
\label{019}
\end{eqnarray}
The last line of this model functional displays the two possible types of couplings of
the DP processes, namely the linear coupling with constant $\sigma $ and the bilinear
coupling with constant $f$. It is easily demonstrated that the renormalization is
complete by noting that the dynamic functional ${\cal J}
_{m}$ describes the renormalizable model of \cite{Ja97} in the case $\sigma
=0$. For $\sigma \neq 0$ it remains to determine the purely multiplicative
renormalization of the relevant operator $\tilde{s}_{2}s_{1}$. Explicitly our arguments
are as follows: For $\sigma =0$, perturbation theory generates only the
ultraviolet-divergent vertex functions $\Gamma _{\tilde{s}
_{i},s_{i}}$, $\Gamma _{\tilde{s}_{i},s_{i},s_{i}}$ with $i=1,2$, and $
\Gamma _{\tilde{s}_{2},s_{1},s_{2}}$, where only the latter depends on $f$.
All the other possible ultraviolet-divergent vertex functions are zero and can only be
generated by insertions of the composed field $\tilde{s}_{2}s_{1}$. Such insertions
make either the vertex functions superficially convergent or produces the
logarithmically divergent $\Gamma _{\tilde{s}_{2},s_{1};\left(
\tilde{s}_{2}s_{1}\right) }$. The logarithmic primitive divergency of the
latter vertex function is eliminated by multiplying $Z_{\sigma }$ defined
via the renormalized composed field by $\left[ \tilde{s}_{2}s_{1}\right]
_{r}=$ $Z_{\sigma }\tilde{s}_{2}s_{1}=Z_{\sigma }Z^{-1}\mathaccent"7017{\
\tilde{s}_{2}}\mathaccent"7017s_{1}$.

We know from our considerations above that the stable one-loop order fixed point
belongs, up to an $\alpha $-transformation, to the minimal model $ {\cal J}_{m}$
(\ref{019}) with the additional constraint $f=g$. In \cite {Ja97} it was shown that
this equality holds at the fixed point in two-loop order also. Thus one may suspect
that $Z_{f}=Z_{g}$ to all orders of the loop expansion provided $f=g$. We will now show
that this conjecture is correct. Moreover we show that $Z_{\sigma }=Z_{\tau }$.

Of course the minimal model ${\cal J}_{m}$ (\ref{019}) is a special case of the
complete model ${\cal J}_{c}$ (\ref{04}) with $f_{1}=f_{1}^{\prime }=f_{2}^{\prime }=0$
and $f_{2}=f$. A comparison between both renormalized dynamic functionals shows that we
can set the coun\-ter\-terms $A$ and $Z_{b}b$ to zero safely in the minimal case. In
addition we find $Y_{1}=Y_{2}=0$ and $ Z_{2}=Z_{f}$. After an $\alpha $-transformation
(\ref{07},\ref{09},\ref{010}) we obtain new coupling constants
\begin{eqnarray}
\bar{\sigma} &=&\sigma +\alpha \left( \tau _{1}-\tau _{2}\right) \;,
\nonumber \\
\bar{f}_{1} &=&2\alpha f+\alpha \left( \alpha -1\right) g\;,\qquad \bar{f}
_{2}=f+\alpha g\;,  \nonumber \\
\bar{f}_{1}^{\prime } &=&\alpha \left( \alpha +1\right) g\;,\qquad\qquad\quad \bar{f}
_{2}^{\prime }=-\alpha g\;.  \label{020}
\end{eqnarray}
and the additive renormalizations
\begin{eqnarray}
\bar{Y}_{1} &=&-\bar{Y}_{2}=\alpha \left( Z_{\tau }-Z_{\sigma }\right) \;,
\nonumber \\
\bar{A} &=&A=\bar{Z}_{b}b=Z_{b}b=0\;.  \label{021}
\end{eqnarray}
Now we demand rapidity reversal as an additional symmetry by equating $\bar{ f
}_{i}=\bar{f}_{i}^{\prime }$. This gives rise to the two equations $\alpha
=-1/2$, $f=g$ (there exists a second solution with $\alpha =0$, $f=0$ that
corresponds to the bilinear uncoupled case and the instable fixed line of
\cite{THH98}). In this rapidity reversal symmetric case the new coupling
constants are given by
\begin{equation}
\bar{f}_{1}=\bar{f}_{1}^{\prime }=-g/4\;,\qquad \bar{f}_{2}=\bar{f}
_{2}^{\prime }=g/2\;,  \label{022}
\end{equation}
and we have from Eq.$\ $(\ref{021}) $\bar{Y}_{1}=-\bar{Y}_{2}=\left( Z_{\sigma
}-Z_{\tau }\right) /2$. Otherwise we know from Eq.\ (\ref{013}) that
$\bar{Y}_{1}=\bar{Y}_{2}$ if rapidity reversal invariance holds. Therefore both
additive renormalizations must be zero and from $\bar{ Y }_{1}=\bar{Y}_{2}=0$ we obtain
the equality
\begin{equation}
Z_{\sigma }=Z_{\tau }\;.  \label{023}
\end{equation}
Thus, $\sigma $ and the $\tau _{i}$ renormalize in the same way and the
crossover exponent $\Phi $ defined by the scaled variable $\sigma /|\tau
_{i}|^{\Phi }$ is given simply by
\begin{equation}
\Phi =1\;  \label{024}
\end{equation}
to all orders of perturbation theory. The other equations (\ref{010},\ref
{013}) yield
\begin{equation}
\bar{Z}_{1}=\bar{Z}_{1}^{\prime }=\bar{Z}_{2}=\bar{Z}_{2}^{\prime
}=Z_{f}=Z_{g}\;  \label{025}
\end{equation}
as one would expect on grounds of consistency.

The discussion of the fixed-point solutions of the complete model ${\cal J}
_{c}$ (\ref{04}) has shown to first order in an $\varepsilon $-expansion
that the stable fixed-point functional belongs to the class of restricted models
described by the minimal functional ${\cal J}_{m}$ up to an $\alpha $
-transformation with the identification $f=g$. Thus the fixed point has a
higher symmetry (rapidity reversal) to this order. This higher symmetry is preserved
under renormalization. It can easily be demonstrated by induction (see appendix B of
\cite{Ja95} of an analogous case) that rapidity reversal symmetry then holds at the
fixed point to all orders. Hence all universal properties of the coupled DP processes
may be calculated from the minimal model ${\cal J}_{m}$ (\ref{019}) with identified
coupling constants $f=g$ and renormalizations $Z_{\sigma }=Z_{\tau },Z_{f}=Z_{g}$. By
virtue of an $
\alpha $-transformation the functional ${\cal J}_{m}$ is equivalent by to
the rapidity reversal invariant functional
\begin{eqnarray}
{\cal J} &=&\int dtd^{d}x\,\lambda   \nonumber \\ \times\!\!\!\!\!\!\!&&
\biggl\{\tilde{s}_{1}\Bigl[Z\lambda ^{-1}\partial_{t}+Z_{\tau }\tau _{1}-Z_{\lambda
}\nabla ^{2}+\frac{g}{2}Z_{g}^{\,}\left( s_{1}-\tilde{s }_{1}\right) \Bigr]s_{1}
\nonumber \\ &&+\tilde{s}_{2}\Bigl[Z\lambda ^{-1}\partial_{t}+Z_{\tau }\tau
_{2}-Z_{\lambda }\nabla ^{2}+\frac{g}{2}Z_{g}^{\,}\left( s_{2}-\tilde{s}
_{2}\right) \Bigr]s_{2}  \nonumber \\
&&-\tilde{s}_{2}\Bigl[Z_{\tau }\bar{\sigma}+Z_{g}\frac{g}{8}\left(
s_{1}-4s_{2}+4\tilde{s}_{1}-\tilde{s}_{2}\right) \Bigr]s_{1}\biggr\}\;
\label{026}
\end{eqnarray}
where $\bar{\sigma}=\sigma -\left( \tau _{1}-\tau _{2}\right) /2$. Note
again that this model solely requires the renormalization constants of the
one-species Gribov process that are known to second order in $u$. We cite
the results of the two-loop calculation \cite{Ja81}
\begin{eqnarray}
Z &=&1+\frac{u}{4\varepsilon }+\frac{u^{2}}{32\varepsilon }\left( \frac{7}{
\varepsilon }-3+\frac{9}{2}\ln \frac{4}{3}\right) +O\left( u^{3}\right) \;,
\nonumber \\
Z_{\lambda } &=&1+\frac{u}{8\varepsilon }+\frac{u^{2}}{128\varepsilon }
\left( \frac{13}{\varepsilon }-\frac{31}{4}+\frac{35}{2}\ln \frac{4}{3}
\right) +O\left( u^{3}\right) \;,  \nonumber \\
Z_{\tau } &=&1+\frac{u}{2\varepsilon }+\frac{u^{2}}{2\varepsilon }\left(
\frac{1}{\varepsilon }-\frac{5}{16}\right) +O\left( u^{3}\right) \;,
\nonumber \\
Z_{u} &=&Z_{g}^{\,2}=1+2\frac{u}{\varepsilon }+\frac{u^{2}}{2\varepsilon }
\left( \frac{7}{\varepsilon }-\frac{7}{4}\right) +O\left( u^{3}\right) \;.
\label{027}
\end{eqnarray}

Computations based on the minimal dynamic functional ${\cal J}_{m}$ (\ref{019}) are
much easier to perform as calculations using the complete model ${\cal J}
_{c}$ (\ref{04}). Thus it may be possible to find the equations of state for
$M_{2}=g\langle s_{2}\rangle $ to second order and check the assumptions
made in \cite{THH98} on the reexponentiation of logarithms to yield the new
order parameter exponent $\beta _{2}$ of that paper (for a calculation of
the equation of state for $M_{1}=g\langle s_{1}\rangle $ to two-loop order
see \cite{JKO98}).

The model ${\cal J}$ (\ref{026}) describes the coupled DP processes near the
multicritical point $\tau _{1}=\tau _{2}=\sigma =0$. What is needed for a thorough
calculation of $\beta _{2}$ is a theory that comprises the limit $\sigma \rightarrow
\infty $. Therefore our considerations here do not solve the problem
addressed in \cite{THH98}, namely the determination of the scaling exponent $
\beta _{2}$ that arises in the scaling $\langle n_{2}\rangle \propto \sigma
^{\beta _{1}-\beta _{2}}\left( \tau _{1c}-\tau _{1}\right) ^{\beta _{2}}$
where species $1$ is in its active phase. T\"{a}uber et al. calculate $\beta
_{2}$ by reexponentiation of logarithms. However their approach relies on
the assumption that simple reexponentiation is possible. To derive such scaling
properties faithfully one indeed has to solve the crossover problem $
\sigma \rightarrow \infty $ which induces (possibly!) a new scaling at
infinity for the correlations of species $2$. Some features of this crossover remind us
of the crossover from special to ordinary behavior in the theory of surface transitions
\cite{Diehl}, with $\sigma $ corresponding to the surface enhancement $c$ and the
species $1$ and $2$ corresponding to the bulk and surface respectively. The crossover
problem of interest here is thus as yet unsolved.

\acknowledgments We thank O.\ Stenull for a critical reading of the
manuscript. This work has been supported in part by the SFB 237 (``Unordnung
und gro\ss e Fluktuationen'') of the Deutsche Forschungsgemeinschaft.

\label{references}

\end{multicols}

\end{document}